\documentclass[prx,twocolumn,superscriptaddress,nopacs]{revtex4-2}

\usepackage{graphicx}
\graphicspath{{figures/}} 
\usepackage{amsmath}
\usepackage{amssymb}
\usepackage{accents}
\usepackage{color}
\usepackage{verbatim}
\usepackage{hyperref}

\usepackage{empheq} 
\definecolor{very-light-gray}{gray}{0.9}

\begin{document}


\title{
Universal scaling near band-tuned metal-insulator phase transitions}

\author{Simone Fratini}
\affiliation{Universit\'e Grenoble Alpes, CNRS, Grenoble INP, Institut Néel, 38000 Grenoble, France}
\author{Sergio Ciuchi}
\affiliation{Dipartimento di Scienze Fisiche e Chimiche, Università dell’Aquila, 67100 Coppito (AQ), Italy}
\affiliation{Istituto dei Sistemi Complessi, CNR, P.le Aldo Moro I-00185 Roma, Italy}
\author{Vladimir Dobrosavljevi\'{c}}
\affiliation{Department of Physics and National High Magnetic Field Laboratory, Florida State University, Tallahassee, FL, USA}
\author{Louk Rademaker}
\affiliation{Department of Quantum Matter Physics, University of Geneva, 1211 Geneva, Switzerland}

\date{\today}

\begin{abstract}
We present a theory for band-tuned metal-insulator transitions based on the Kubo formalism. Such a transition exhibits scaling of the resistivity curves, in the regime where $T\tau >1$ or $\mu \tau>1$, where $\tau$ is the scattering time and $\mu$ the chemical potential.
At the critical value of the chemical potential, the resistivity diverges as a power law, $R_c \sim 1/T$. Consequently, on the metallic side there is a regime with negative $dR/dT$, which is often misinterpreted as insulating. 
We show that scaling and this `fake insulator' regime is observed in a wide range of experimental systems.
In particular, we show that Mooij correlations in high-temperature metals with negative $dR/dT$ can be quantitatively understood with our scaling theory in the presence of $T$-linear scattering.
\end{abstract}

\maketitle




Thanks to the advent of highly tunable `twisted' Van der Waals heterostructures,\cite{Balents.2020,Kennes.2021,Mak.2022} the field of quantum matter physics is in a position to study continuous zero-temperature phase transitions with an unprecedented accuracy. Detailed (and smooth!) experimental results allow a systematic comparison between different theoretical predictions, which is particularly true for continuous {\em metal-to-insulator transitions} (MITs).

Interaction-induced MITs, such as the Mott transitions, display quantum critical behavior, including scaling of the resistivity.\cite{Ghiotto.2021,Li.202109b} A full theoretical understanding of Mott criticality, which would include a precise calculation of the scaling exponents, is still lacking.\cite{Tan.2022} One of the main challenges lies in the fact that an MIT is, in general, {\em not} a transition described by symmetry breaking, which makes it challenging to identify the source of scaling.

Recently, scaling has been observed in a simple {\em band-tuned} MIT in a MoTe$_2$/WSe$_2$ bilayer at full filling of the first valence flat band.\cite{Li.2021nyj} By tuning the displacement field, one can open a band gap to the second valence band. The scaling behavior there has been analysed using a model with disorder and a bosonic 
field,\cite{Tan.2022dde} inspired by earlier work on `Mooij' correlations.\cite{Mooij.1973,Ciuchi.2018} However, the observed scaling can also be interpreted in a much simpler perspective.

From a theoretical viewpoint, calculating the conductivity is notoriously difficult. An exception is the classical Drude formula, $\sigma = \frac{ne^2 \tau}{m}$, which can also be derived with fully quantum-mechanical advanced methods such as the Kubo formula,\cite{Mahan.2000,Coleman.2015p0w}.
A natural question is whether the observed scaling at a metal-insulator transition can be explained with the same set of assumptions that is used to derive Drude theory.

Indeed, in this Letter we show that only a small number of very natural assumptions leads to scaling behavior near a band-tuned MIT. The only assumptions are that the scattering time $\tau$ is large, parametrized by $T \tau >1$ or $\mu \tau>1$ respectively on the insulating and metallic sides of the transition (with $\mu$ the chemical potential measured from the band edges), and that the electron self-energy is local and proportional to the electron density of states. These conditions naturally arise in weakly correlated, weakly disordered metals. With this, the critical resistivity at the MIT is diverging as $R_c(T) \sim 1/T$, in contrast to oft-cited picture that the critical resistivity curve is independent of temperature. We derive an explicit scaling form, showing that in the scaling regime the resistivity is given by a universal $R(T,\mu) = R_c(T) f(\mu/T)$. Contrary to the physics of universality at continuous phase transitions, the scaling of the resistivity breaks down very close to the MIT. 





\begin{figure}
	\centering
	\includegraphics[width=\columnwidth]{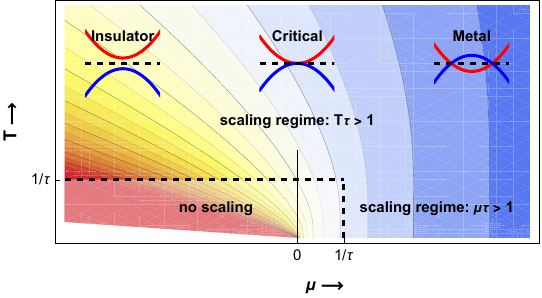}
	\caption{In a band-tuned metal-insulator transition (MIT), the system changes from having overlapping valence (blue) and conduction (red) bands in the metallic side (right) to having a gap on the insulating side (left). 
    The tuning parameter is the chemical potential $\mu$.
    When either $T\tau > 1$ or $\mu \tau > 1$, the resistivity (in shades on the background) can be described by a scaling form, as shown in Fig.~\ref{Fig:MITscaling}. This scaling relation breaks down very close to the transition, where localization and interaction effects will change the picture.}
	\label{Fig:SimpleModel}
\end{figure}

\begin{figure*}
	\centering
	\includegraphics[width=\textwidth]{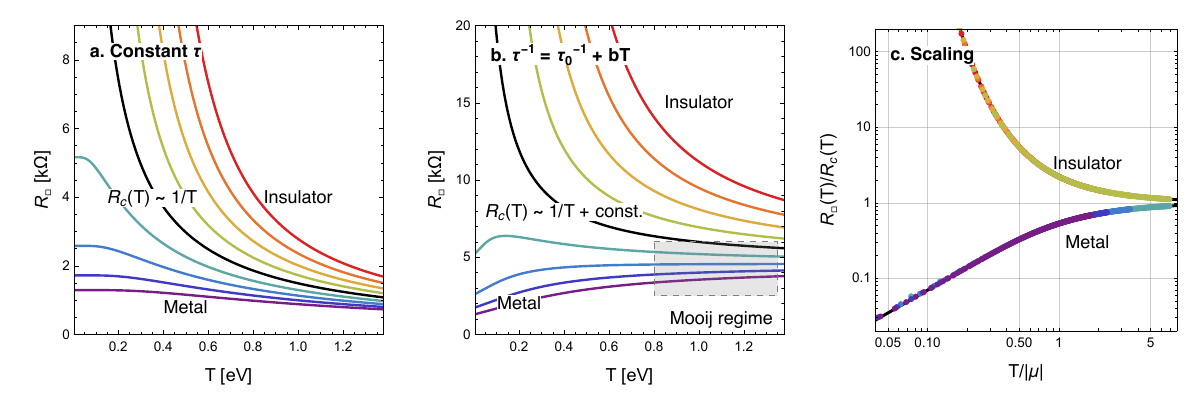}
	\caption{Theoretical resistance curves close to a band-tuned metal-insulator transition.
    {\bf a.} Resistance calculated using Eq.~\eqref{Eq:FiniteTEFconductivity}, for a constant scattering time $\tau = 25$ eV$^{-1}$, and chemical potential $\mu$ ranging from $-0.8$ to $+0.8$ eV. 
    The resistance at the critical point $\mu =0$ diverges as $R_c(T) \sim 1/T$. 
    On the metallic side, the resistance decreases as a function of temperature (a `fake' insulator), whereas on the insulating side the resistance is activated.
    {\bf b.} Resistance for a temperature-dependent scattering rate $\tau^{-1} = \tau_0^{-1} + b T$ with $\tau_0 = 25$ eV$^{-1}$ and $b = 0.1$. On the metallic side, there is a transition from a positive $dR/dT$ to a negative $dR/dT$. At high temperatures, this gives rise to Mooij correlations (see Fig.~\ref{fig:TCRexperiments}).
    {\bf c.} When $T\tau >1$ or $\mu \tau>1$, the resistance curves follow a simple scaling law $R (T, \mu) = R_c(T) f(\mu/T)$. This can be verified by plotting $R/R_c$ versus $T/|\mu|$. All data points collapse onto one of the two curves, associated with either metallic or insulating behavior.}
	\label{Fig:MITscaling}
\end{figure*}

{\em Band-tuned MIT -- } Consider a weakly interacting electron system described by a band-structure.
The system is metallic if there is a nonzero density of charge carriers, characterized by a nonzero chemical potential $\mu$. 
The system is an insulator if there is a gap towards exciting charge carriers. By continuously changing the bandstructure we can induce a band-tuned MIT.
This can be achieved with pressure, displacement field, or even due to spontaneous symmetry breaking such as ferromagnetic polarization.
Without loss of generality, the dispersion at a band edge is parabolic, with the dispersion set by $\xi_{\bf k} = \frac{k^2}{2m} - \mu$ where $m$ is the effective mass. With this notation, $\mu >0$ corresponds to the metal, $\mu < 0$ to an insulator, and $\mu = 0$ is the critical point. The chemical potential $\mu$ is thus the {\em tuning parameter} of the MIT, as shown in Fig.~\ref{Fig:SimpleModel}.

In general, the conductivity is determined by {\em disorder}, {\em electron-electron} interactions and {\em electron-phonon} coupling. 
Nonzero resistivity from electron-electron interactions requires Umklapp scattering, which becomes asymptotically irrelevant at low carrier densities (though there might be nontrivial vertex corrections)\cite{Mu.2022}.
Similarly, at zero temperature there is no thermal occupation of phonons, and therefore no electron-phonon contribution to the resistivity.
The zero-temperature behavior of a band-tuned MIT is therefore completely dominated by disorder.
In principle strong disorder might push the system into Anderson insulation. However, in $d=2,3$ it is considered that the combination of weak disorder and weak interactions generally precludes true localization \cite{Evers.2008,Abanin.2019k1s,Punnoose.2005,Castellani.1984}. Moreover, even in the absence of interactions, quantum corrections to the conductivity are not relevant in the regimes  $\mu \tau >1$ and $T\tau>1$ considered here, and will therefore be neglected throughout this work.

\begin{figure*}
	\centering
	\includegraphics[width=\textwidth]{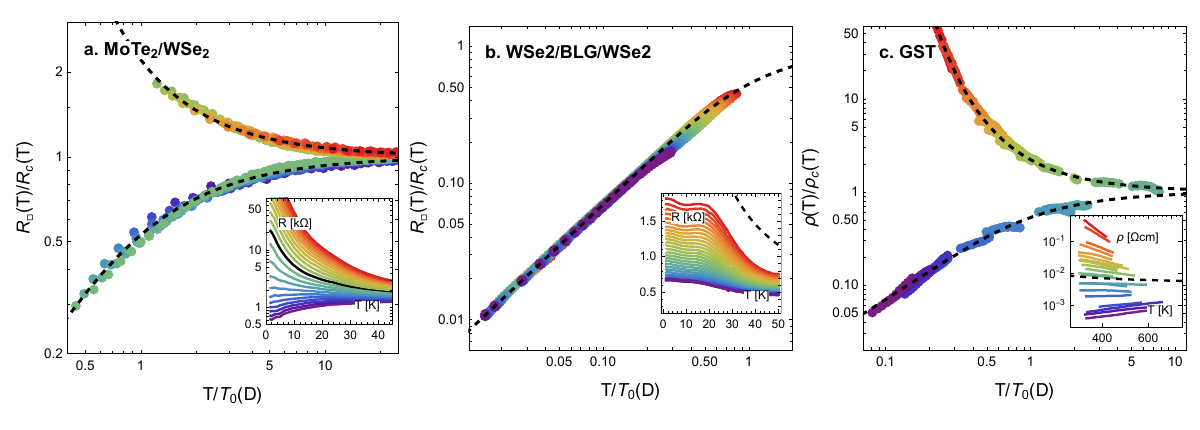}
	\caption{ 
 Scaling near the band-tuned MIT is observed in a range of materials. Here, we apply our scaling analysis to three material systems\cite{sm}: ({\bf a}) the moir\'e heterobilayer MoTe$_2$/WSe$_2$\cite{Li.202109b}, ({\bf b}) the heterostructure WSe$_2$/bilayer graphene/WSe$_2$\cite{Kedves.2023}, and ({\bf c}) GST amorphous phase change materials \cite{Siegrist.2011}. The measured resistivities are shown in insets. In panels a and c we see a genuine MIT, with data collapse on both an insulating and conducting branch. The theoretical scaling curve of Eq.~\eqref{eq:ScalingForm} is shown as a dashed black line, and shows remarkable agreement with the experimental results.}
	\label{Fig:MITexperiments}
\end{figure*}

{\em Conductivity -- } With these natural assumptions, the conductivity close to the MIT is calculated using the Kubo formula for local self-energies \cite{Mahan.2000, Georges.1996}, which reads for a single band in $d$ dimensions, per spin species,
\begin{equation}
    \sigma_{xx} = 
        \pi
        \int d\xi
            \Phi_x (\xi)
        \int dz
        A^2(\xi, z)
        \left( -f' (z)\right),
    \label{Eq:KuboNoVertex}
\end{equation}
where $A(\xi,z)$  is the one-particle spectral function and $f$ is the Fermi function (see \cite{sm}, Sec.~\ref{Sec:Kubo}). The entire momentum-dependence is included in a {\em transport function}
   $ \Phi_x (\xi) = \int \frac{d^dp}{(2\pi)^d}
        j^2_x ({\bf p}) \, \delta( \xi - \xi_{\bf p} )$.
The transport function itself displays universal behavior in the vicinity of a band-tuned MIT: given the parabolic band dispersion, the current operator equals ${\bf j}({\bf p}) = \frac{e}{m} {\bf p}$. Consequently the transport function reads
\begin{equation}
    \Phi_x (\xi) = \frac{2 e^2}{dm} \left( \xi + \mu \right) N(\xi),
\end{equation}
where $N(\xi)$ is the non-interacting density of states, $N(\xi) =\int \frac{d^dp}{(2\pi)^d}  \delta( \xi - \xi_{\bf p} )$. 
Assuming a constant, energy-independent scattering rate $\tau$ in $d=2$, the imaginary part of the self-energy is $\mathrm{Im} \Sigma(z) = -\Theta(z+\mu) (2\tau)^{-1}$. This scattering time is typically of the order $\tau \sim 10^{-12} - 10^{-14}$ s $\sim 10 - 10^3$ eV$^{-1}$. When $\mu \tau > 1$ or $T \tau > 1$, the Kubo formula radically simplifies, and we find
\begin{equation}
    \sigma (T, \mu) = \frac{e^2}{h} 
		\tau T \log \left[ 1 + e^{\mu/T}\right].
	\label{Eq:FiniteTEFconductivity}
\end{equation}
This is our central result for the conductivity close to the band-tuned MIT. Surprisingly, it contrasts a few commonly held convictions on metal-insulator transitions. First, at the critical point, the conductivity is {\em linear} in temperature, $\sigma_c(T) \equiv \sigma(0,T) = \frac{e^2}{h} \tau T \log 2$, rather than temperature-independent. 
Furthermore, on the metallic side of the transition $(\mu > 0)$, the temperature derivative of the resistivity can be negative: a `fake insulator' regime that is commonly misinterpreted as insulating.
Furthermore, Eq.~\eqref{Eq:FiniteTEFconductivity} satisfies a universal scaling form 
\begin{align}
    \sigma(\mu ,T ) &= \sigma_c(T) F(\mu/T),\label{eq:ScalingForm}
\end{align}
which allows the collapse of many resistivity curves onto a simple scaling function $F (x)= \log_2 \left[ 1 + e^x \right]$. 
The theoretical resistance curves near the band-tuned MIT, including the scaling properties, are shown in Fig.~\ref{Fig:MITscaling}.

Hidden in plain view is the fact that Eq.~\eqref{Eq:FiniteTEFconductivity} is, at zero temperature on the metallic side, equivalent to Drude theory. Explicitly, in $d=2$, $\sigma(\mu) = \frac{e^2}{h} \tau \mu = \frac{ne^2 \tau}{m}$. 
In fact, the $T=0$ limit of Eq. (\ref{Eq:KuboNoVertex}) yields $\sigma=\Phi(E_F) \tau$ with  $\Phi(E_F)=ne^2/m$ in any dimension $d$.


At finite temperature the scaling regime persists, even with a temperature-dependent scattering time $\tau(T)$, provided that $\tau^{-1}$ is still proportional to the density of states. When $\tau$ is temperature-independent, in fact, {\em all} resistivity curves on the metallic side are `fake insulators' with $d \rho /dT \le 0$ (cf. Fig. \ref{Fig:MITscaling}). Only when the scattering rate increases with temperature, for example from electron-phonon interactions shown in Fig.~\ref{Fig:MITscaling}b or from Umklapp scattering, we find traditional metallic behavior with $d\rho/dT>0$. In this case, inside the metallic regime there exists a point where the temperature-derivative of the resistivity $d\rho/dT$ changes sign. We will discuss universal properties around this point later in the context of Mooij correlations.\cite{Mooij.1973,Lee.1985}

It is important to emphasize that the scaling form of Eq. (\ref{eq:ScalingForm}) is limited to regions {\em not too close} to the transition. This limitation is similar to the one proposed by Mott-Ioffe-Regel (MIR) \cite{Hussey.2004}. A common formulation of the  MIR limit in metals is $k_F \ell \sim 1$ where $\ell$ is the mean-free path. This can be rewritten as $\mu \tau \sim 1$; we therefore find that, upon approaching the transition from the metallic side, the scaling hypothesis breaks down precisely at the MIR boundary.
What happens close to the transition is non-universal, and depending on model parameters one can find various different violations of scaling (see \cite{sm}, Sec.~\ref{Sec:Nonuniversal}).




{\em Band-tuned MIT in moir\'e bilayers -- } We are now in a position to verify our universal scaling result of Eq.~\eqref{Eq:FiniteTEFconductivity} in experimental results on real physical systems. 
Inspired by the recent developments in moir\'e materials, let us first focus on the MIT in MoTe$_2$/WSe$_2$ at full filling of the first valence flat band ($f=2$).\cite{Li.2021nyj} By tuning the perpendicular displacement field, a gap is opened up, yielding a band-tuned MIT. In Fig.~\ref{Fig:MITexperiments} we fit the observed resistance curves as a function of displacement field using our theory. Indeed, the critical resistance diverges as $R_c \sim 1/T$, and the resistance curves obey scaling. As shown in Fig.~\ref{Fig:MITexperiments}a, the scaling curve itself quantitatively matches the analytical form derived in Eq.~\eqref{Eq:FiniteTEFconductivity}.
A similar scaling plot for these data has been reported in Ref. \cite{Tan.2022dde}, inspired by earlier work in
Ref. \cite{Ciuchi.2018}, which describes disorder-induced polaron formation when the chemical potential is {\em far} from the band edges. 
When the chemical potential approaches the band edges, the theory of Ref. \cite{Tan.2022dde} reduces to the simpler theory presented here, where polaronic effects are irrelevant.

There are many claims of MITs in graphene-based Moir\'e materials, that upon closer inspection seem to exhibit "fake insulator" behavior. Consider, for example, the WSe$_2$/bilayer graphene (BLG)/WSe$_2$ heterostructure measured in Ref.~\cite{Kedves.2023}. At filling $\nu=0$, the resistivity turns up at low temperatures reminiscent of an insulating gap. However, at around $T=20 K$, the resistivity seems to saturate, to a displacement-field dependent value. The absence of a true diverging resistance at low temperature suggests that these systems retain a nonzero density of charge carriers, either from a band overlap or induced by potential inhomogeneities that are common in graphene systems. 
Indeed, when performing the scaling analysis, we can collapse all the curves of this system to the {\em metallic} branch of our scaling form, as shown in Fig.~\ref{Fig:MITexperiments}b.

{\em Disordered metallic alloys -- }
While Eq.~\eqref{Eq:FiniteTEFconductivity} was derived for $d=2$ and weak disorder scattering, it is in fact far more universal. Often a momentum-independent self-energy arises through the equation $\Sigma(z)=s^2 G(z)$, for example in iterated schemes such as the self-consistent Born approximation for disorder scattering or electron-phonon scattering in the adiabatic limit. Here $s$ is a (possibly temperature dependent) parameter quantifying the scattering process. Under this scheme, the inverse scattering time is in weak coupling proportional to the density of states ${\rm Im} \Sigma(z) \propto N(z) \sim \mathrm{Im} G(z)$. This leads to a conductivity of the form $\sigma(T)=\frac{e^2}{d m s^2} T \log \left[ 1 + e^{\mu/T} \right]$ in general dimensions $d$, consistent with Eq.~\eqref{Eq:FiniteTEFconductivity}.

The universal scaling is indeed also observed in three-dimensional compounds away from the weak disorder limit. In particular, we look at GST\cite{Siegrist.2011}, a phase-change compound where the annealing history affects the effective number of charge carriers.\cite{Zhang.2012} Here, at high temperatures, a smooth evolution from positive $dR/dT$ to negative $dR/dT$ is observed depending on the precise composition and history of the sample. Since the main effect of these compositional changes is in fact a shift of the chemical potential, we show in Fig.~\ref{Fig:MITexperiments}c that the experimental data on GST can be accurately described by our scaling theory.

\begin{figure}[t]
    \centering
    \includegraphics[width=\columnwidth]{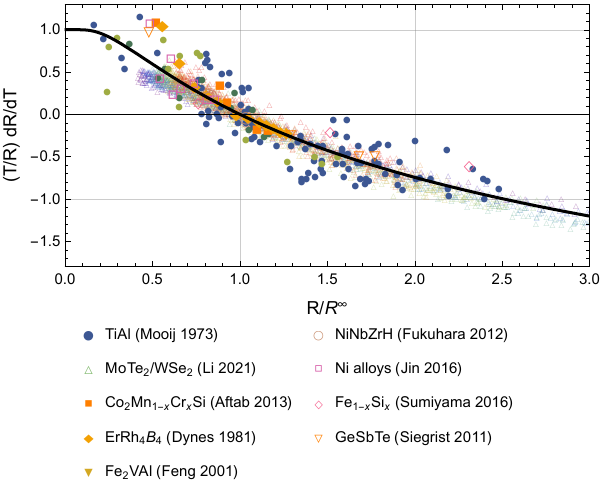}
    \caption{The dimensionless temperature coefficient of the resistance versus the dimensionless resistance, for a variety of materials\cite{Mooij.1973,Li.202109b,Aftab.2013,Dynes.1981ternary,Feng.2001,Fukuhara.2012,Jin.2016,Sumiyama.2016,Siegrist.2011} (for details see \cite{sm}), compared to our theoretical result of Eq.~\eqref{Eq:MooijAnalytical} (solid black line). For the experimental data the only fitting parameter is $R^\infty$, the limit of the critical resistance at high temperature. We find an excellent agreement of the experimental Mooij correlations and our theory.}
    \label{fig:TCRexperiments}
\end{figure}

{\em Mooij correlations -- } 
Universal scaling implies the existence of a `{\em fake insulator}' regime: a metal characterized by a (dimensionless) negative temperature coefficient of the resistance $\alpha = (T/R) dR/dT<0$.
Historically, the observation of a negative $\alpha$ various disordered metals, including binary alloys (Ni$_x$Cr$_{1-x}$, Ti$_x$Al$_{1-x}$, Fe$_x$Si$_{1-x}$, etc.)\cite{Mooij.1973,Ciuchi.2018} was considered a `high temperature anomaly'.\cite{Lee.1985} In a seminal paper, Mooij\cite{Mooij.1973} discovered a correlation between the temperature coefficient $\alpha$ and the resistivity $\rho$ itself. There is currently no consensus on the origin of these Mooij correlations, though they have been interpreted in terms of quantum localization corrections to the conductivity \cite{Tsuei.1986,Lee.1985} or the disorder-driven formation of polarons\cite{Ciuchi.2018}.

Interestingly, the scaling theory proposed in this Letter allows to quantitatively describe Mooij correlations. To do so, we assume that at high temperature the scattering time $\tau$ is  linear in $T$:
\begin{equation}
    \tau^{-1} = \tau_0^{-1} + b T.
\end{equation}
This form occurs in many metals, where $b$ is either proportional to the electron-phonon coupling strength, or a more complex, "Planckian" quantum scattering.\cite{Bruin.2013}
With this assumption, the critical curve becomes flat at high temperature, $R_c(T) \rightarrow R^\infty \propto b$. This allows us to introduce a dimensionless resistivity $R/R^\infty$. 
By taking the derivative of the scaling relation Eq.~\eqref{Eq:FiniteTEFconductivity}, and inverting it with respect to the tuning parameter $\mu$ at a fixed temperature $T$, we find that the temperature coefficient $\alpha$ only depends on $R/R^\infty$,
\begin{equation}
    \alpha (R) = 
        \frac{R}{R^\infty}
        \left( 1 - 2^{R^\infty/R} \right)
        \log_2 \left[ 2^{R^\infty/R} -1 \right].
    \label{Eq:MooijAnalytical}
\end{equation}
In Fig.~\ref{fig:TCRexperiments} we compare our analytical result with the original data presented by Mooij\cite{Mooij.1973} and those collected in \cite{Ciuchi.2018}, finding a good agreement between the experimental results on binary alloys and Eq.~\eqref{Eq:MooijAnalytical}. The recent data on Moir\'e bilayers shows an even more striking {\em quantitative} equivalence between the resistivity data in the high temperature range $T = 26 - 60$ K: without a fitting parameter the experimental results of Ref.~\cite{Li.202109b} match Eq.~\eqref{Eq:MooijAnalytical}.




{\em Outlook --} In this Letter we have shown that a simple theory of conductivity predicts universal scaling near band-tuned MITs consistent with experimental results in a wide range of materials, from recent Moir\'e materials to decades old data on binary alloys.

The predicted scaling regime does not extend arbitrarily close to the MIT: when $T\tau<1$ and $\mu \tau<1$ deviations from or a full breakdown of scaling can appear.
Note that the difference between scaling close and further away from the transition has been discussed in Ref.~\cite{Tan.2022}.
The scaling described in this Letter is thus {\em not} due to the divergence of a length scale, and is {\em not} related to Landau order parameters, the renormalization group, or any other theory of universality in symmetry-breaking (quantum) phase transitions. The universal behavior of resistivity scaling near the MIT throughout many materials is just the consequence of a generic weakly interacting electrons with weak disorder, in spirit similar to the stability of the Fermi liquid.
The properties of Anderson and weak localization as well as Wigner crystallization and the Mott MIT\cite{Tan.2022} are phenomena that, on the other hand, are outside the scaling regime discussed here.
It is an interesting open question whether the scaling described in this Letter can extend, under certain conditions, arbitrarily close to the MIT, thus connecting to the standard theoretical framework of continuous phase transitions.\cite{Mahmoudian.2015}


{\em Acknowledgements -- } We thank Jie Shan, Kin Fai Mak, P\'eter Makk and B\'alint Szentp\'eteri for sharing their experimental data. We thank Yuting Tan, Christophe Berthod, and Giacomo Morpurgo for fruitful discussions. LR is funded by the Swiss National Science Foundation by Starting Grant TMSGI2\_211296.
SC is funded by the European Union - NextGenerationEU under the Italian Ministry of University and Research (MUR) National Innovation Ecosystem grant ECS00000041 - VITALITY - CUP E13C22001060006.

\bibliographystyle{myapsrev}
\bibliography{libraryMIT}


\clearpage
\onecolumngrid
\appendix
\begin{center}
    \Large{Supplementary information}    
\end{center}

\section{Derivation of conductivity using Kubo formula}
\label{Sec:Kubo}

In the absence of vertex corrections the Kubo formula for conductivity reads
\begin{equation}
	\sigma_{xx} = 
		\pi
		\int \frac{d^dp}{(2\pi)^d}
		\int dz
		\;
		A({\bf p}, z)
		j_x ({\bf p})
		A({\bf p}, z)
		j_x ({\bf p})
		\left( -f'(z) \right).
	\label{Eq:KuboNoVertex}
\end{equation}
where ${\bf j} ({\bf p}) = \frac{e {\bf p}}{m}$ is the current operator, $f(z) = [ e^{z/T}+1]^{-1}$ is the Fermi function, and $A({\bf p}, z)$ is the spectral function
\begin{equation}
	A({\bf p},z) = 
	\frac{1}{\pi} \frac{ - \mathrm{Im} \Sigma({\bf p},z)}
		{ (z - \xi_{\bf p} - \mathrm{Re} \Sigma({\bf p}, z))^2 + (\mathrm{Im} \Sigma({\bf p},z))^2}.
\end{equation}
Here $\Sigma({\bf p},z)$ is the electron self-energy. When the self-energy is independent of momentum, we can replace the momentum integral by an integral over dispersion $\xi$ and introduce the transport function as discussed in the main text,
\begin{eqnarray}
    \Phi_x (\xi) &=& \int \frac{d^dp}{(2\pi)^d}
        j^2_x ({\bf p}) \, \delta( \xi - \xi_{\bf p} ) \\
   &=& \frac{2 e^2}{dm} \left( \xi + \mu \right) N(\xi).
\end{eqnarray}
so that the Kubo formula reads
\begin{equation}
      \sigma_{xx} = 
        \pi
        \int d\xi
            \Phi_x (\xi)
        \int dz
        A^2(\xi, z)
        \left( -f' (z)\right)
\end{equation}
Now we include the real part of the self-energy in a renormalization of the bare dispersion; and assume the imaginary part of the self-energy is ${\rm Im} \Sigma(z) = -\frac{1}{2\tau} \Theta(z + \mu)$. In this case, relevant for the novel moir\'e systems, we can exactly integrate over $\xi$ in $d=2$ dimensions,
\begin{eqnarray}
    \sigma (T, \mu)& =& 
    \frac{e^2}{2 \pi h}
    \int_{-\mu} dz \left( -f'(z)  \right)
    \left[ 1 + 2 \tau (z + \mu)
			\left(\frac{\pi}{2} + \mathrm{atan} \left( 2 \tau (z + \mu) \right) \right)
			\right], \label{eq:sigma2d}
\end{eqnarray}
The leading order term in the limit where $(\mu + z)\tau$ is large, becomes
\begin{eqnarray}
    \sigma (T, \mu)& =& 
    \frac{e^2}{h} \tau
    \int_{-\mu} dz \left( -f'(z)  \right)
    (z + \mu)
    = \frac{e^2}{h} \tau T \log \left[ 1 + e^{\mu/T} \right]
\end{eqnarray}
which is the central result of the main text.

\section{Nonuniversal behavior close to the transition}
\label{Sec:Nonuniversal}

The scaling ansatz is in general {\em not valid} arbitrarily close to the transition. Explicit breakdown of scaling can be seen in theoretical models even within the weak-coupling Kubo formula without vertex corrections (meaning: even if we ignore Anderson localization, Mott localization, Wigner crystallization, and percolation).

The scaling ansatz $\sigma(\mu ,T ) = \sigma_c(T) F(\mu/T)$ in the limit of $T \rightarrow 0$ implies, with $\sigma_c(T) \sim T$, that $\sigma(T=0) \propto \mu$. This is trivially true for the Drude formula in $d=2$, whereas in $d=3$ it requires ${\rm Im} \Sigma \propto \sqrt{\mu}$. The {\em breakdown} of scaling close to the transition can thus be inferred from having {\em nonlinear} behavior of $\sigma$ as a function of $\mu$ at zero temperature.

In 2nd order perturbation theory with disorder we obtain ${\rm Im} \Sigma = - \frac{1}{2\tau_0}$. The zero-temperature conductivity is thus, for $\mu>0$,
\begin{eqnarray}
    \sigma (\mu, T=0) = \frac{e^2}{2 \pi h} \left[ 1 + 2\mu \tau_0 \left( 
    \frac{\pi}{2} + \arctan 2 \mu \tau_0 \right) \right]
\end{eqnarray}
and $\sigma=0$ on the insulating side. This result features a jump at $\mu=0$ of magnitude $\Delta \sigma = \frac{e^2}{2\pi}$, breaking the scaling ansatz.

This jump persists even when the density of states do not have a discontinuity at the band edge. In the self-consistent Born approximation (SCBA) the self-energy is given by
    $\Sigma (z) = \gamma G(z)$
where $G(z)$ is the fully dressed Green's function and $\gamma$ the strength of the disorder. Note that within this scheme, the density of states $A(z) = -\frac{1}{\gamma \pi} {\rm Im} \Sigma(z)$, which is now continuous at the band edge. Nevertheless, there is still a (nonuniversal) jump in the conductivity.

In $d=3$, integrating over momenta in the Kubo formula yields the zero-temperature conductivity for $\mu>0$,
\begin{equation}
    \sigma(\mu, T=0) = 
    \frac{\sqrt{m} 
    \left(\text{Im$\Sigma (\mu)$}^2+\mu ^2\right)^{3/4} 
    \left(\sqrt{\frac{|\text{Im$\Sigma (\mu)$}|}{\sqrt{\text{Im$\Sigma (\mu)$}^2+\mu ^2}}+1}+\sqrt{1-\frac{|\text{Im$\Sigma (\mu)$}|}{\sqrt{\text{Im$\Sigma (\mu)$}^2+\mu ^2}}}\right)^3}
    {24 \sqrt{2} \pi ^2 \text{Im$\Sigma (\mu)$}}
\end{equation}
Note that indeed when $\mu > {\rm Im} \Sigma$ this reduces to the Drude formula $\sigma = \frac{e^2 \sqrt{2m} \mu^{3/2} \tau}{3 \pi^2} = \frac{e^2 n \tau}{m}$. In the main text we assumed for scaling to hold $\tau^{-1}$ to be proportional to the density of states, which is proportional to $\sqrt{\mu}$ near the band edge. This results in a conductivity proportional to $1/\sqrt{\tau}$ close to the transition, violating the scaling ansatz

\begin{figure*}
	\centering
	\includegraphics[width=\textwidth]{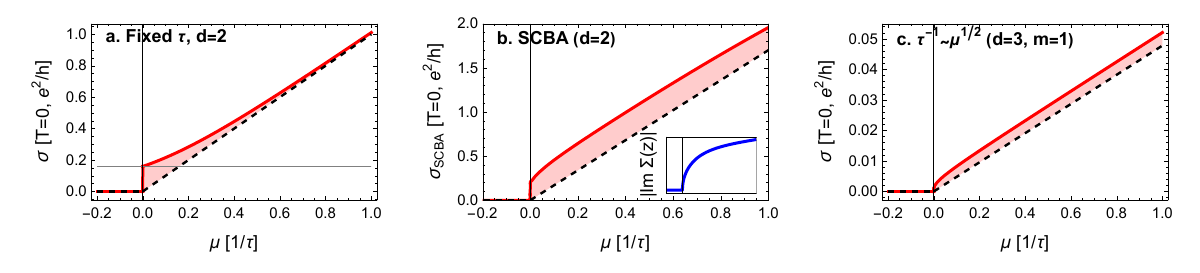}
	\caption{
The theoretical zero-temperature conductivity displays a breakdown of scaling (in black dashed lines) near the MIT, in $d=2$ for fixed $\tau=1$ ({\bf a.}), the self-consistent Born approximation ({\bf b.}), and in $d=3$ for $\tau^{-1} = \sqrt{\mu}$ ({\bf c.}). The red shaded area indicates the deviation from scaling. The inset in {\bf b.} shows the imaginary part of the self-energy for the SCBA with $\gamma =1$ in the same energy range as the main figure.}
	\label{Fig:JumpSM}
\end{figure*}

All three cases are visualized in Fig.~\ref{Fig:JumpSM}.

\section{Analysis of the experimental data: Scaling}

\begin{figure*}
	\centering
	\includegraphics[width=\textwidth]{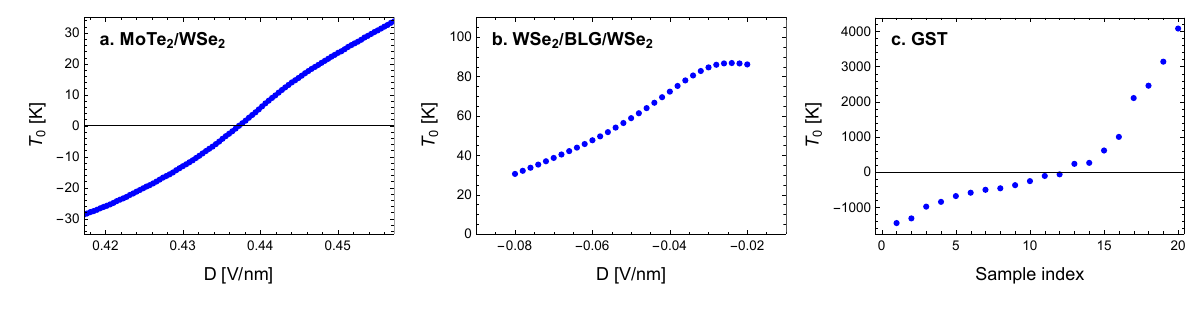}
	\caption{
In the experimental scaling plots in the main text, we divide for each curve the temperature by a scaling parameter $T_0$. Here we show the value of $T_0$ as a function of tuning parameter for the three considered experimental systems. 
{\bf a.} The $\nu=2$ metal-insulator transition observed in MoTe$_2$/WSe$_2$\cite{Li.202109b}, the tuning parameter is the vertical displacement field $D$ measured in V/nm. We identify the critical $D_c$ as where the extrapolated $\sigma(T=0)$ vanishes. 
{\bf b.} In WSe$_2$/BLG/WSe$_2$, the tuning parameter is the vertical displacement field. We could not identify a critical value of $D_c$. Instead, we found $R_c(T)$ by optimizing the data collapse of all the experimental curves. Note that we restricted our dataset to a regime of $D$ where the change in resistance is monotonic. The value $T_0$ can be identified as a chemical potential for conducting charge carriers. These are possibly the result of potential fluctuations present in bilayer graphene.
{\bf c.} The different samples of GST each have a different annealing history and composition of Ge, Sb and Te, as outlined in Ref.~\cite{Siegrist.2011}. Here we ordered the data sets based on the scaling parameter $T_0$ ranging from insulating ($T_0<0$) to metal ($T_0>0$).}
	\label{Fig:MITexperimentsSM}
\end{figure*}

In order to perform scaling, we used publicly available resistance/resistivity data from Refs.~\cite{Li.202109b,Kedves.2023,Siegrist.2011}. For each data set, we first identified the critical curve of the form
\begin{equation}
    R_c (T) = \frac{1}{\tau_0 T} + b
    \label{Eq:CriticalRc}
\end{equation}
where $\tau_0$ is the disorder-induced scattering time and $b$ is some linear-$T$ scattering rate that can come from phonons or general `Planckian' dissipation. Written like Eq.~\eqref{Eq:CriticalRc}, these two parameters have units that depend on the dimension ($d=2,3$). For the three experimental systems, we used the following parameters for the critical curve:

\begin{center}
\begin{tabular}{c|c|c}
    System & $\tau_0$ & $b$ \\
    \hline
     MoTe$_2$/WSe$_2$\cite{Li.202109b} & 192 k$\Omega$/K & 2.72 k$\Omega$ \\
     WSe$_2$/BLG/WSe$_2$\cite{Kedves.2023} & 84.1 k$\Omega$/K & -- \\
     GeSbTe\cite{Siegrist.2011} & 1.35 $\Omega$cm/K & 0.003 $\Omega$cm
\end{tabular}
\end{center}

We rescaled the temperature with a parameter $T_0$ (which can be interpreted as the chemical potential $\mu$) as a function of tuning parameter to achieve a collapse of all resistance/resistivity curves. The relevant values of $T_0$ are shown in Fig.~\ref{Fig:MITexperimentsSM}.

For MoTe$_2$/WSe$_2$ and WSe$_2$/BLG/WSe$_2$ (BLG stands for bilayer graphene), the tuning parameter is a vertical displacement field. The data reported are at filling $\nu=-2$ and $\nu=0$, respectively, relative to charge neutrality. 

The label GST refers to a collection of different compounds GeSb$_4$Te$_7$, GeSb$_2$Te$_4$,  Ge$_2$Sb$_2$Te$_5$, and Ge$_3$Sb$_2$Te$_6$. Different resistivity curves correspond to those four materials each annealed at a different temperature, as outlined in the Supplementary Information of Ref.~\cite{Siegrist.2011}.

\section{Breakdown of scaling}

\begin{figure}
	\centering
	\includegraphics[width=0.6\textwidth]{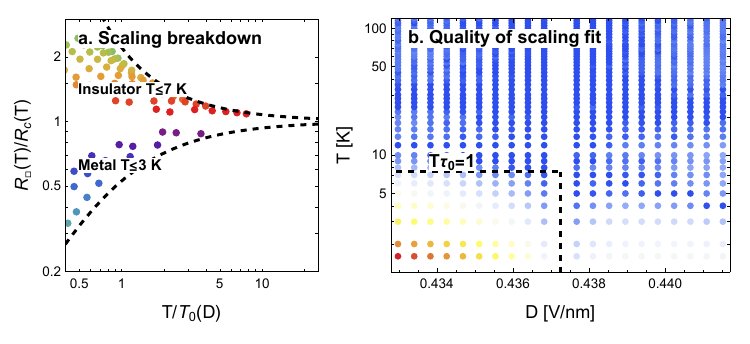}
	\caption{At low temperatures ($T\tau \leq 1$) we expect a breakdown of the scaling regime. This is indeed observed in MoTe$_2$/WSe$_2$\cite{Li.202109b}. {\bf a.} Here we show only the low-temperature data ($T \leq 3$ or $7$ K, respectively on the insulator or metal side of the MIT), scaled with the same $T_0$ and $R_c$ as is used for the figure in the main text. The experimental data points are clearly not on the theoretical scaling curve (dashed line). {\bf b.} We can quantify the deviation from scaling by calculating the distance-squared between the experimental data points and the theoretical curve. Here we show in colorscale (blue equals good fit, red is bad fit) the quality of the scaling ansatz as a function of displacement field and temperature. Using the estimate of $\tau_0$, we show with a dashed line where one would expect scaling to break down.}
	\label{Fig:ScalingBreakdownSM}
\end{figure}

Note that in the case of MoTe$_2$/WSe$_2$, as expected, the scaling breaks down at low temperatures. Details of this are shown in Fig.~\ref{Fig:ScalingBreakdownSM}. Note that the breakdown of scaling in the epxerimental system is consistent with the theoretical picture of Fig.~\ref{Fig:SimpleModel}. 

\section{Analysis of the experimental data: Mooij correlations}

The data presented in Fig.~4 of the main manuscript is collected from a variety of sources. The analysis to arrive at the dimensionless temperature coefficient of the resistivity $\alpha$ is for most materials\cite{Aftab.2013,Dynes.1981ternary,Feng.2001,Fukuhara.2012,Jin.2016,Sumiyama.2016,Siegrist.2011} based on our earlier analysis presented in Ref.~\cite{Ciuchi.2018}.

The TiAl data presented is from Fig.~6 of Ref.~\cite{Mooij.1973}. The three different sets of data presented there are shown in Fig.~4 of the main manuscript with filled circles of different colors.

The Mooij correlations for MoTe$_2$/WSe$_2$ are based on the same data analyzed for the scaling of Fig.~3\cite{Li.202109b}, limited to the temperature range $T = 26$ -- 63 K, and the range of displacement fields $D = 0.476$ -- 0.408 V/nm. The temperature derivative of the resistance is calculated with a two-point forward finite difference. The only fitting parameter is $R^\infty = 1.44 $ k$\Omega$ is obtained by collapsing the data for different displacement fields onto the same curve. In Fig.~4 of the main manuscript, the data for MoTe$_2$ is presented with an empty upward triangle where the different colors represent the different displacement fields.

\end{document}